\documentclass[graybox]{svmult}

\usepackage[numbers]{natbib}
\usepackage{type1cm}        % activate if the above 3 fonts are
\usepackage{makeidx}         % allows index generation
\usepackage{graphicx}        % standard LaTeX graphics tool
\usepackage{multicol}        % used for the two-column index
\usepackage[bottom]{footmisc}% places footnotes at page bottom

\usepackage{newtxtext}       % 
\usepackage[varvw]{newtxmath}       % selects Times Roman as basic font

\newcommand{\Gaia}{\textit{Gaia}}
\newcommand{\hst}{\textit{HST}}
\newcommand{\jwst}{\textit{JWST}}
\newcommand{\Ho}{$H_0$}

\def\mWH{$m_H^W$}

\makeindex             % used for the subject index
\begin{document}

\title*{On Cepheid Distances in the \Ho\ measurement}
\author{Richard I. Anderson}
\institute{Richard I. Anderson \at Institute of Physics, \'Ecole Polytechnique F\'ed\'erale de Lausanne (EPFL), Observatoire de Sauverny, Chemin Pegasi 51, 1290 Versoix, Switzerland, \email{richard.anderson@epfl.ch}}
% \and Name of Second Author \at Name, Address of Institute \email{name@email.address}}
\maketitle

\abstract*{Classical Cepheids were the first stellar standard candles and have played a crucial role for astronomical distance measurements ever since the discovery of the Leavitt law (period-luminosity relation). Enormous improvements in distance accuracy have been achieved since Hertzsprung's first application of Leavitt's law to measure the distance to the Small Magellanic Cloud in 1913, notably in very recent years thanks to a large data set of highly accurate space astrometry from the ESA mission \Gaia. Complemented by homogeneous space photometry, Cepheids enable the most accurate distance estimates to galaxies hosting type-Ia supernovae up to approximately 70 Mpc distant. Here, I review the history of Cepheid distance measurements, open questions on the side of stellar astrophysics, and recent studies seeking to quantify and mitigate systematics with a view to further improve the accuracy on the Hubble constant. For example, the recently launched James Webb Space Telescope will enhance precision due to 4x lower sensitivity to source blending in crowded regions and greater sensitivity in dust-insensitive infrared bands. Future 30m-class telescopes could in principle further improve Cepheid distance measurements towards the Hubble flow, if technical challenges related to a continuously evolving instrument can be overcome.}

\abstract{Classical Cepheids were the first stellar standard candles and have played a crucial role for astronomical distance measurements ever since the discovery of the Leavitt law (period-luminosity relation). Enormous improvements in distance accuracy have been achieved since Hertzsprung's first application of Leavitt's law to measure the distance to the Small Magellanic Cloud in 1913, notably in very recent years thanks to a large data set of highly accurate space astrometry from the ESA mission \Gaia. Complemented by homogeneous space photometry, Cepheids enable the most accurate distance estimates to galaxies hosting type-Ia supernovae up to approximately 70 Mpc distant. Here, I review the history of Cepheid distance measurements, open questions on the side of stellar astrophysics, and recent studies seeking to quantify and mitigate systematics with a view to further improve the accuracy on the Hubble constant. For example, the recently launched James Webb Space Telescope will enhance precision due to 4x lower sensitivity to source blending in crowded regions and greater sensitivity in dust-insensitive infrared bands. Future 30m-class telescopes could in principle further improve Cepheid distance measurements towards the Hubble flow, if technical challenges related to a continuously evolving instrument can be overcome.}

\section{Cepheids: From variable stars to cosmic yardsticks}\label{sec:history}

The variability of classical Cepheids (henceforth for simplicity: Cepheids, cf. Sec.\,\ref{sec:Cepheids}) was first discovered by visual obserations in the 18th century by John Goodricke and Edward Piggott \citep{Goodricke1786}. Friedrich Argelander began systematically surveying stars on both hemispheres in the 1850s, and Charles Pickering pushed for systematic observations of variable stars towards the end of the 19th century. Pickering hired female assistants (the so-called Harvard ``Computers'') to process the observations collected \citep{sobel2016glass}. The photographic plates were astrometrically aligned and variable stars identified by blinking observations of the same sky area observed at different epochs. Henrietta Leavitt thus conducted an unprecedented census of 1777 variable stars in the Small Magellanic Cloud \citep{Leavitt1908}. 

In 1912, Pickering reported on Leavitt's groundbreaking discovery based on 25 classical Cepheids, that the apparent magnitude $m$ of 25 SMC Cepheids correlated well with the logarithm of their period of variability $P$, both near minimum and maximum light \citep{Leavitt1912}. Since all SMC stars were assumed to reside at a common distance, it was clear that the observed relation between brightness and period must be an intrinsic property of the stars. This period-luminosity relation (PLR) is nowadays referred to as Leavitt's law (LL). 

Just one year later, Ejnar Hertzsprung \citep{Hertzsprung1913} made the first attempt to measure distance using Leavitt's discovery. He estimated the absolute magnitude of Cepheids at a period near $6.6$\,d  using secular parallax and, in turn, determined the distance modulus of the SMC Cepheids by comparing their apparent magnitude at the same period to that of the MW Cepheids. Hertzsprung's work laid the foundation for the distance ladder in use today. Despite concluding on an incorrect distance ($2000$
lightyears), he convincingly demonstrated that the SMC is external to the MW based on a comparison of the scale height of stars in the Solar Neighborhood. Thus, Cepheid distances already informed the (extragalactic) distance scale before their variability mechanism was understood \cite{Shapley1914,Eddington1917,Wesselink1946}, and well before the Great Debate of 1920, Hubble's works on M33 \citep{Hubble1926}, as well as the subsequent discovery of the Hubble-Lema\^itre law \citep{Lemaitre1927,Hubble1929}. 

\subsection{The Hubble constant measured using the distance ladder}

The relation between Hubble's constant, \Ho, and $D_L$ is given by the Friedmann equation, developed here to second order using the Robertson-Walker metric:
\begin{equation}
D_L = \frac{cz}{H_0} \left[ 1 + \frac{1}{2}(1-q_0)z - O(z^2) +  \ldots \right] \ ,
\label{eq:FLRW}
\end{equation}
where $q_0 \approx -0.55$ is the \textit{deceleration} parameter, whose observed negative value implies the Universe's accelerated expansion \citep{riess1998,Perlmutter1999}. Clearly, the wish to measure \Ho\ accruately is intimately tied to measuring accurate distances.

Eq.\,\ref{eq:FLRW} requires the measured redshifts to be purely cosmological, that is, any motions due to graviational attraction (large scale flows, peculiar motions, etc.) must be subtracted or negligible, so that \Ho\ is best measured in the Hubble flow at distances of $90 - 600$\,Mpc ($0.01 < z < 0.0233$) \citep{Riess2016}. Such distances are accessible thanks to the extremely luminous type-Ia supernovae (SNeIa, cf. Vincenzi this volume) that enable $\sim 3\%$ relative distance estimates. However, since SNeIa are very rare, no SNIa has as yet been observed in a galaxy whose distance is known by geometrical means. SNeIa thus require external calibration by other means, such as Cepheids, via the \emph{distance ladder (DL)}. 

\begin{figure}[t]
\centering
\includegraphics[width=1\textwidth]{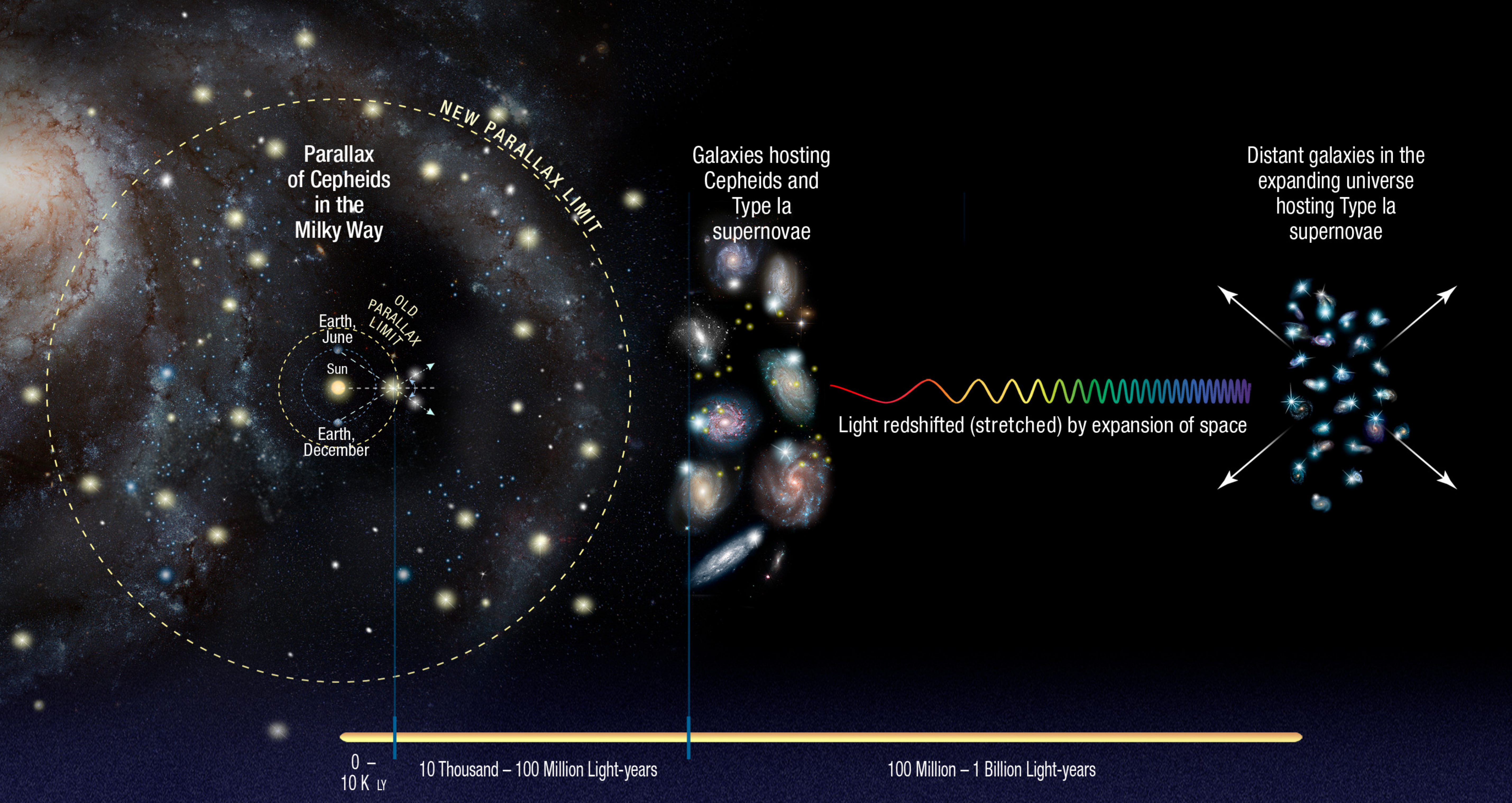}
\caption{The extragalactic distance ladder uses a two-rung approach for measuring \Ho. Geometric distances calibrate stellar standard candles for determining luminosity distances, which in turn calibrate the luminosity of SNeIa that trace the Hubble flow. Credit: NASA, ESA, A.~Feild (STScI), and A.~Riess (STScI/JHU)\label{fig:DL}}
\end{figure}

The modern DL consists of three rungs as illustrated in Fig.\,\ref{fig:DL}. On the first rung, Cepheids are calibrated as standard candles (SCs) using distances measured by geometric methods. On the second rung, Cepheids are used to determine distances to galaxies (SN-hosts) where at least one SNIa has been observed, out about $80$\,Mpc \citep{Riess2022H0}.  On the last rung, SNeIa map cosmic expansion in the Hubble flow. 
\label{sec:modern}

Seen the other way around, the DL connects the SNeIa distance-redshift relation to an angular scale provided by geometry using Cepheids as the intermediary. By further connecting the Hubble diagram to the early Universe's angular scale (the cosmological sound horizon) via the cosmic expansion history ($H(z)$) using baryon acoustic oscillations, the DL enables a crucial cosmological end-to-end test involving two angular scales at opposite ends of the cosmos\hbox{---}trigonometric parallaxes and the quantum fluctuations of the early Universe \citep{Riess2020nature}.

\Ho\ measurements using a DL have a rich history (cf. Tully, this volume), involving much controversy and disagreements. The \hst\ Key Project to measure the Hubble constant was designed to resolved the ongoing disputes and for the first time achieved a $10\%$ measurement of \Ho\ \citep{Freedman2001}. 

In the late 2000's, the SH0ES (Supernova \Ho\ for the Equation of State) project was launched to measure \Ho\ in order to determine the nature of Dark Energy using a differential DL built on the principle of data homogeneity \citep{Riess2009a,Riess2009}. In its most recent form, the SH0ES DL is implemented as a least squares problem that minimizes \citep{Riess2022H0}:
\begin{equation}
\chi^2 = (y - Lq)^T C^{-1} (y - Lq)  \ ,
\label{eq:DL}
\end{equation}
with the data vector $y$, the design matrix $L$ that encodes the relevant equations, such as the LL, distance moduli, etc., the covariance matrix $C$, and the best-fit parameter vector $q$. Covariance among observables can be directly considered by quantifying the corresponding elements of $C$, and very significant amount of covariance information is already being considered, notably among background corrections of Cepheids in SN host galaxies \citep{Scolnic2022,Brout2022,Riess2022H0}. Using 3445 degrees of freedom, 5 best fit parameters are determined by the SH0ES DL (see Sec.\,\ref{sec:cal} for definitions): the LL slope $\beta$ ($b_W$), metallicity effect $\gamma$ ($Z_W$), the fiducial absolute magnitude of a 10-day Cepheid in the {\it HST} NIR Wesenheit magnitude $M_{H,1}^W$, the absolute $B-$band magnitude of SNeIa, $M_B^0$, and Hubble's constant $5 \log_{10}{H_0}$. Alternative symbols used in literature are listed in parenthesis. 

The matrix formalism is both simple and accurate, and allows straightforward investigation of analysis variants, notably to assess the systematic effects of modeling choices or dataset selection. Nevertheless, the results from the matrix inversion are cross-checked with computationally intensive Markov chain Monte Carlo (MCMC) simulations that allow inspection of marginalized posterior distributions and correlations among fit parameters. TRGB distances have also been included in this procedure and can help to further improve precision. Including the latest results based on cluster Cepheids, this DL yields $H_0= 73.15 \pm 0.97 \, \mathrm{km\,s^{-1}\,Mpc^{-1}}$ \citep{Riess2022clusters}.

The specific construction of the SH0ES DL, with the simultaneous fit to all observables resembles a tightly controlled \emph{experiment} that seeks to maximize the benefits of purely differential flux comparisons to leverage statistical precision while clearly quantifying the systematic uncertainties involved in calibrating the absolute luminosity scale. In this spirit, the SH0ES DL emphasized the use of homogeneous and very well calibrated Hubble Space Telescope (\hst) photometry. The greatly improved spatial resolution and enhanced sensitivity of the James Webb Space Telecope (\jwst) is increasingly replacing the near-infrared (NIR) photometry observed using \hst\ WFC3/IR \citep{Riess2023crowd}. Careful cross-calibration between \hst\ and \jwst\ NIR photometry is thus required to ensure direct compatibility. As a result of its specific construction, any modifications to the experimental setup must consider consequences across the entire DL in order to ensure apples-to-apples comparisons.

\subsection{On standard candles, calibration, and standardization}\label{sec:cal}

Astronomical standard candles (SCs) are objects that fulfill at least two conditions: a) their intrinsic brightness (absolute magnitude $M$) can be calibrated in a given photometric band, or combinations of several bands, and b) they can be faithfully recognized in different observational contexts, notably in different galaxies. SCs are crucial to astronomy because they allow to determine \emph{luminosity distances}, $D_L$ (cf. Eq.\,\ref{eq:DL}), by exploiting the fact that the observed flux $F$ decreases with distance squared, that is, $F \propto L / D_L^2$. In magnitude space, this becomes the distance modulus equation $\mu = m - M - A = 5\log_{10}{D_L} + 25$, with $D_L$ in Mpc and extinction $A$. As noted by Hertzsprung, using standard candles for distance determination presumes that the absolute magnitude established in one context, e.g., in the Milky Way, $M_{\mathrm{cal}}$, is representative of the absolute magnitude of sources seen in other contexts, such as SN host galaxies, $M_0$. Assuming this correspondence, distance is determined by comparing the observed (apparent) magnitude to the calibrated absolute magnitude, $\mu = m_{\mathrm{obs}} - M_{\mathrm{cal}} - A$. 

\emph{Calibrating} standard candles to establish $M_{\mathrm{cal}}$ requires observing them at known distance, $M_{\mathrm{cal}} = m - \mu - A$. Galaxies where this is possible are frequently referred to as \emph{anchors}, and three anchors are of particular importance for Cepheids: The Milky Way (MW), the Large Magellanic Cloud (LMC), and the water megamaser-host galaxy NGC\,4258. Despite an available geometric distance estimate of the SMC, complications related to its elongated line-of-sight orientation have thus limited its usefulness for measuring \Ho\ at the current levels of precision. Interestingly, the distance measurements used for all three anchors are systematically very different. In the MW, trigonometric parallaxes of SCs, notably from the European Space Agency \Gaia\ mission (cf. Lindegren, this volume, and references therein) \citep{Prusti2016gaia,Brown2021gaiaedr3,GaiaDR3summary}, set the current gold standard. In particular, parallaxes of open star clusters that host Cepheids have provided the most accurate Cepheid calibration to date ($0.9\%$ in distance) \citep{CruzReyes2023} thanks to better understood systematics of the fainter, non-variable cluster stars, and the statistical gain of averaging over many cluster members. The typical uncertainty of such cluster parallaxes is $7\,\mu$arcsec, three times better than for individual field Cepheids, so that one cluster Cepheid carries as much weight as 9 field Cepheids \citep{Riess2022clusters}.  

The geometric distance measurements of the LMC (detached eclipsing binaries) and NGC4258 (Keplerian disk model of observed megamaser features observed using VLBI) rely on very different systematics, including instrumentation used and modeling assumptions, rendering them truly independent checks of the calibration based on parallax. The close agreement among the Cepheid calibrations based on these anchors is therefore a strong attestation to the accuracy of the anchor distances as well as the consistency among Cepheid observations spanning an enormous contrast range from approximately $8$th to $28$th magnitude (8 dex in flux).  

\emph{Standardization} refers to the process of mitigating differences that disturb the equivalence between $M_{\mathrm{cal}}$ and $M_0$. All stellar standard candles (including Cepheids, SNeIa, and the TRGB method) require standardization to provide distances to better than a few percent.

As Leavitt showed more than a century ago, Cepheids differ very significantly in luminosity depending on their pulsation period (by about 3 mag/dex of period), which ranges from a few days to more than 100 days. Cepheid distances are thus obtained by comparing \emph{fiducial} magnitudes. The most commonly adopted form of the Leavitt law of Cepheids is 
\begin{equation}
M (P) = \alpha + \beta \cdot \log{P/P_0} + \gamma \cdot \mathrm{[Fe/H]} \ ,
\label{eq:LL}
\end{equation}
with $\alpha$ the fiducial absolute magnitude of a Solar metallicity Cepheid at the pivot period $P_0$ (also often referred to as the zero-point), $\beta$ the LL slope, and $\gamma$ the metallicity term that allows to correct the impact of chemical composition using iron (also: oxygen) abundances, cf. below. Importantly, $\alpha$, $\beta$, and $\gamma$ generally depend on the photometric passband (or combinations thereof) used.

The fiducial absolute Cepheid magnitude in the SH0ES DL, $M_{H,1}^W$, represents a $10$\,d Cepheid at the center of the instability strip in a specific combination of photometric passbands (cf. Sec.\,\ref{sec:Wesenheit}). The commonly used pivot period of $10$\,d is motivated by literature reports of LL non-linearity \citep{Bhardwaj2016} and the period-amplitude diagram of Cepheids \citep{Klagyivik2009}, and corresponds to an intermediate period among Cepheids in anchor galaxies, although Cepheids at greater distance tend to have increasingly longer periods. $\beta$ and $\gamma$ are determined in the global DL fit using all Cepheids in anchor and SN host galaxies simultaneously.

At present, there is no clearly demonstrated need to consider LL non-linearity when using Cepheids to measure distances, and allowing for two different slopes near the pivot period does not significantly improve the fit. Moreover, the astrophysical motivation for using non-linear LLs is limited, since stellar models of single chemical composition yield log-linear LLs across a large range of periods \citep[e.g.][]{Anderson2016rot,DeSomma2021PA}. However, stellar evolution models do predict a small dependence of chemical composition on LL slope \citep{Anderson2016rot}. Since the effect is rather small, it has not yet been confidently detected \citep{Ripepi2020}. LL slope changes due to metallicity are furthermore particularly small in the infrared, rendering the SH0ES Wesenheit formalism very insensitive to them. 

It seems that observed LL non-linearity arises in a context-specific way from an uneven sampling of the period-color relation, which is caused by the finite width of the classical instability strip. Multi-phase PL relations \citep{Ngeow2012} compare LLs at different pulsation periods and have demonstrated broken PL relations. However, Cepheid magnitudes used for measuring distances correspond to the average flux of Cepheids, and cannot be directly compared to PL relations observed at specific phases. The coincidence of the reported LL breaks near $10$\,d strongly suggest an origin in the \cite{Hertzsprung1926} progression. 

Two further considerations are in order concerning LL non-linearity. Firstly, \Ho\ bias due to LL non-linearity arises only in case of systematic differences among the distributions of Cepheid periods in anchor and SN-host galaxies, and this effect is of order $\lesssim 0.15\%$ (the author \& Bastian Lengen, priv. comm.). 
Secondly, variants of the SH0ES DL that include a break period of $10\,d$ are presented alongside the baseline analysis. These variants also indicate a very insignificant effect on \Ho, while not being preferred according to the change in $\chi^2$.

Aside from pulsation period, the most common parameter used for standardization considers metallicity differences \citep{Kennicutt1998,Rizzi2007,Freedman2011,Breuval2022}. For Cepheids, most studies assume a global offset to the LL using a linear dependence on chemical composition usually quantified via oxygen or iron abundances relative to Solar composition, that is, $\Delta M_{\mathrm{met}} = M_0 - M_{\mathrm{cal}} = \gamma \cdot \Delta[O/H]$. Determining $\gamma$ is challenging because the effect is rather small ($\sim -0.25$\,mag/dex) \citep{Riess2022H0,Breuval2022}, and metallicity differences among most Cepheids used in the DL are typically less than $0.5-0.6$\,dex. Oxygen abundances estimated in SN-host galaxies fall within a rather narrow range that is fully contained within the abundance ranges of Cepheids in the Milky Way, LMC, and the SMC \citep{Riess2022H0}. As a result, the uncertainty in the metallicity effect does not significantly affect \Ho. However, metallicity can be a limiting factor for determining distances to individual Cepheids, e.g., to map Galactic structure, or for determining \Gaia's parallax offsets \citep{Molinaro2023}.

Standardization is best done using quantities than can be directly observed in the specific sources of interest, such as individual elemental abundances derived from Cepheid spectra. Such comparisons are presently limited to MW and LMC Cepheids, although future 30m class telescopes will significantly increase the ability to observe individual stellar spectra in other nearby galaxies. However, the majority of SN-host galaxies will remain off-limits to individual star spectroscopy. At such distances, equivalent information based on the closest proxys is typically used, such as oxygen abundances of H\,II regions derived by the strong line method \citep{Zaritsky1994,Bresolin2011}, or abundances of blue supergiants  \citep{Kudritzki2008,Bresolin2022}. Ensuring the compatibility of proxy information used for standardization is thus crucial to obtaining unbiased distance estimates. For the realm where this has been feasible, such agreement has been shown, for example among HII regions in the Milky Way  and Cepheid abundances determined by optical spectra \citep{Riess2022H0}.

\subsubsection{A sidenote on TRGB standardization\label{sec:TRGB}}
As all SCs, the TRGB method \citep{Lee1993} is also subject to standardization. However, metallicity corrections are complicated by the statistical nature of the measurement ($m_{\mathrm{TRGB}}$ is measured based on a sample, not for each star individually) and the diverse population of giant stars composed of different ages and chemical composition. As a result, there is less agreement over how population differences should be treated, with some arguing for color-based metallicity corrections \citep{Rizzi2007} and others assuming that differences among the bluest (interpreted to be the oldest) stars will be negligible \citep{Freedman2021}. 

Ubiquitous small amplitude pulsations observed in luminous red gaints near the tip \citep{Ita2002,Eyer2002} provide useful addition information that may resolve this deadlock. \cite{Anderson2023} recently showed that all stars near the TRGB feature are variable, and that the variabilty of red giants provides crucial information about the composition of the diverse red giant population. Comparison between the Large and Small Magellanic Clouds (Koblischke \& Anderson, in prep.) reveals that the PL sequences of small amplitude red giants in the OGLE (Optical Gravitational Lensing Experiment) catalog of variable stars (OSARGs) are systematically shifted to shorter period in the SMC. This is indicative of a metallicity effect that causes lower-metallicity stars to be more compact, increasing the average density of stars, and thus reducing their periods, since $P \propto 1/\sqrt{\rho}$. Thus, OSARG-like variability can be used to standardize metallicity differences for the TRGB method, provided such low levels of variability (amplitudes of order $0.01-0.04$\,mag) can be detected.

Another form of standardization is required for TRGB distances due to methodological choices. The CATs (Comparative Analysis of TRGBs) team \cite{Wu2023,Li2023,Scolnic2023} demonstrated a significant tip-contrast relation (TCR) for TRGB measurements, whereby fields observed in the same (very distant) galaxy differed in TRGB magnitude according to the ratio of the number of stars that are fainter and brighter ($R = N_+ / N_-$) than the TRGB feature. Standardization to a fiducial value of $R=4$ was thus proposed to ensure consistency. Since TCR standardization does not depend on distance, the effect mainly increases the scatter of TRGB distances to SN host galaxies, rather than biasing \Ho. Meanwhile, \cite{Anderson2023} showed that the TCR is a methodological artifact introduced by weighting the edge detection response (Sobel filter output) in an effort to assign statistical significance to it \citep{Madore2009} and that unweighted edge detection responses do not exhibit a significant TCR. Moreover, simulations showed that unweighted EDRs yield less biased TRGB magnitudes than weighted EDRs \citep{Anderson2023}. Since $R$ is defined \emph{after} determining $m_{\mathrm{TRGB}}$, TCR corrections are slightly circular, and unweighted EDRs should be the preferred choice of measuring $m_{\mathrm{TRGB}}$. Additional details on the TRGB method are presented in Beaton (this volume), and \cite{Beaton2018review} provides a detailed account of other methodological aspects of TRGB measurements that must be considered to obtain consistent results.

\section{Cepheids and where to find them -- observational considerations\label{sec:obs}\label{sec:Cepheids}}

\subsection{On the usefulness of Cepheids in the DL\label{sec:CepheidsDL}}

Classical Cepheids afford the best precision for measuring distances thanks to the fortuitous combination of a) high luminosity, b) highly regular variability, c) comparatively straightforward identification thanks to large (optical) amplitudes, d) and relatively high prevalence in star forming galaxies. Of course, Cepheids also come with their own observational challenges and are not suitable for every context. For example, Cepheids in highly inclined galaxies suffer from enhanced crowding and reddening issues. In such cases, the Tip of the Red Giant Branch (TRGB) method can offer a suitable alternative \citep{Lee1993}.

An important advantage of Cepheids is that they are observed specifically and individually. This differs notably from the detection of the TRGB feature, which is ascribed to an astrophysically extremely short-lived phenomenon and which is always measured using an inhomogeneous and contaminated (e.g., by AGB stars) sample of stars. Galaxies where multiple types of SCs can be observed thus provide cross-checks essential for understanding systematic uncertainties while allowing to cross-calibrate multiple SC types, e.g., to increase the number of SN host galaxies in the DL. The following reviews observational aspects of Cepheids and any known issues involving their calibration or use as SCs.

\subsection{Cepheid measurements used in the DL}\label{sec:variability}

\subsubsection{Cepheid types, variability, and classification}
This chapter uses the term ``Cepheid'' synonymous with classical (type-I) Cepheids that belong to the young population I, and whose prototype is $\delta$~Cephei. However, it is important to note that other Cepheid types exist, notably the type-II (subgroups W~Vir, BL~Her, and RV~Tau) and anomalous Cepheids that belong to the old population II and exhibit relatively similar light variability, at least in terms of timescales and amplitudes. Distinguishing these groups is crucial to ensure distance accuracy and indeed led to a major revision of the distance scale by Walter Baade \citep{Baade1956}.

Classical Cepheids are, first and foremost, identified by their variability. Additional context, such as Galactic latitude for MW stars, or the position within a galaxy, usefully separates between young and old stars. However, misclassification between type-I and type-II Cepheids is relatively common if parallax information is not reliable or available. For stars at a common distance, PL sequences are extremely powerful at distinguishing among different types, notably in the Magellanic Clouds, where many PL sequences have been identified \citep{Soszynski2015}. Since classical Cepheids are easily identified by their high luminosity, large number, and period range, there is little room for confusion among Cepheid types in SN host galaxies. 

Where available, multi-band time-series data are helpful to establish temperature variations that distinguish chromatic pulations from usually achromatic eclipse-related phenomena. Spectroscopy can provide certainty by allowing to measure effective temperature variations, asymmetric line shapes (a telltale of pulsations), RV variations, as well as surface gravity and abundances.

\Gaia's magnitude limited survey has provided the largest homogeneous sample of Cepheids classified solely based on (multi-band) light curves, containing 15\,006 Cepheids of all sub-types, including approximately 12\,554 classical Cepheids. These Cepheids belong to galaxies within the Local Group ($ < 1$\,Mpc), including the MW, the Magellanic Clouds, M31, and M33, and nearby dwarfs. In the MW, there are approximately $3500$ classical Cepheids \citep{Pietrukowicz2021,GDR3Cepheids}, whereas nearly $10\,000$ Cepheids are nowadays known in the Magellanic System \citep{Soszynski2015,GDR3Cepheids}.  

The prime targets for \Ho\ are Cepheids that pulsate in the fundamental radial mode (FM Cepheids). These are the most numerous, most regular in terms of their variability, and the longest-period Cepheids. Their detection is further aided by saw-tooth shaped light curves that exhibit a characteristic dependence on pulsation period. This so-called Hertzsprung progression \citep{Hertzsprung1926} provides important evidence of the similarity of nearby and distant Cepheids. FM Cepheids have their highest amplitudes between approximately $5-7$ and $15-30$\,d, and amplitudes decrease for the longest-period Cepheids \citep{Klagyivik2009}. 

Shorter-period Cepheids that pulsate in the first overtone (FO Cepheids) feature sinusoidal light curves of lower amplitudes, and higher overtones feature progressively lower amplitudes that can become challenging to detect and distinguish from other types of variability. Approximate relations between FO and FM pulsation periods \citep{MACHO} are occasionally used to combine samples of FO and FM Cepheids. In these cases, FO Cepheid periods are ``fundamentalized'' using period ratios of FO and FM Cepheids in well observed samples, such as the LMC. However, the metallicity dependence of $P_{\mathrm{FO}}/P_{\mathrm{FM}}$ \citep{Kovtyukh2016} complicates this process and adds systematic uncertainty that is difficult to quantify \citep[e.g.][]{Csoernyei2023M51}. As a result, FO Cepheids are not usually considered in the modern DL.

Besides being more luminous, FM Cepheids also have the advantage of being more regular in their variability than FO Cepheids \citep{Evans2015MOST,Anderson2014,Smolec2016,Sueveges2018a,Sueveges2018b}. However, the very long period Cepheids ($P \gtrsim 25$\,d) exhibit period fluctuations that can be stochastic or (semi-)periodic in nature \citep{Turner2005}, notably causing issues with de-phasing to mean magnitudes due to variable pulsation ephemerides \citep{Csoernyei2022periods}. However, even very fast period changes of $100$\,s/yr ($\dot{P}/P \sim 4 \cdot 10^{-5}$ at $30$\,d, $\Delta \log{P} \sim 6 \cdot 10^{-7}$) are irrelevant for determining distances using LLs.

\subsubsection{Reddening-free Wesenheit magnitudes\label{sec:Wesenheit}}
Near-infrared (NIR) photometry significantly improves the use of Cepheids as standard candles due to the intrinsically reduced sensitivity to extinction, lower amplitudes, and an intrinsically reduced width of the instability strip, which results in lower dispersion. Until recently, the main drawback of NIR photometry for the DL was the reduced spatial resolution of \hst's WFC3/IR channel compared to WFC3/UVIS. However, NIRCAM on board the \jwst\ now provides $4\times$ better spatial resolution than WFC3/IR, with a larger aperture, resulting in much improved Cepheid photometry at large distances \citep{Riess2023crowd}. Unfortunately, the reduced amplitudes of Cepheids in the NIR compared to the optical render Cepheid discovery using \jwst\ not applicable to large samples.

Issues related to dust extinction are commonly mitigated by computing Wesenheit\footnote{The German word ``Wesenheit'' relates to the abstract innate nature (or essence) of an entity, cf. \url{https://www.dwds.de/wb/Wesenheit}.} magnitudes \citep{VanDenBergh1975,Madore1982}, $m^W$, which are reddening-free by construction for a given reddening law \citep[SH0ES uses $R_V=3.3$ from][]{Fitzpatrick1999,Schlafly2011}.  The SH0ES near-IR Wesenheit formalism combines $H-$band (F160W) magnitudes with optical colors, \mWH $= \mathrm{F160W} - R^W\cdot(\mathrm{F555W}-\mathrm{F814W})$, where $R^W = A_H / (A_V - A_I) = 0.386$ set by the reddening law. For Cepheids observed by \Gaia, $R^W_G = A_G / (A_{G_{BP}} - A_{G_{RP}}) \approx 1.91$. Importantly, $R^W$ depends on intrinsic source color when using wide-band photometry due to the shape of the spectral energy distribution incident on the photometric passband \citep{Anderson2022}. The SH0ES approach is thus particularly robust against reddening issues thanks to the combination of intrinsically extinction-insensitive NIR photometry with a rather small color-dependent term. For example, a $10\%$ change in reddening law (from $R_V = 3.3$ to $3.0$) would result in a $0.01 - 0.05$ change in $R_H$ (depending on reddening law), which multiplied by a typical color excess of $E(V-I) \approx 0.6$ would merely change \mWH\ by $\sim 0.03 * 0.6 = 0.02$\,mag, that is, much less than the dispersion in most SN host galaxies, cf. Fig.\,22 in \citep{Riess2022H0}.

Wesenheit magnitudes have the desirable side-effect of narrowing LL dispersion due to the finite width of the instability strip, since $R^W$ happens to be close to the slope of the period-color relation. Of course, Wesenheit magnitudes require knowledge of the reddening law, which can vary among sightlines, and this can be an issue for specific sightlines, such as Cepheids in the SMC, which has been reported to have an abnormal reddening law \citep{Gordon2003}. However, given the thousands of sightlines among Cepheids in anchor galaxies (MW, LMC, NGC4258) and the thousands of sightlines to Cepheids in SN-host galaxies it is highly unlikely for reddening-law related bias to significantly affect \Ho, as was also found empirically \citep{Moertsell2022,Riess2022H0}.

\subsubsection{Cepheids in the Milky Way}\label{sec:MW}

MW Cepheids are particularly important for the DL due to the ability to measure trigonometric parallax \citep{Feast1997}. However, MW Cepheids are subject to several challenges that arise from their location inside the MW's disk, where dust extinction is high and a strong function of distance, and where the LL cannot be recovered using only apparent magnitudes. Observationally, the brightness of Cepheids can be an asset in that high-resolution spectra can be observed \citep{daSilva2022,Ripepi2022,Trentin2023}. However, the same brightness can easily saturate photometric observations of Cepheids, notably from space, using large apertures, and in the infrared. Another practical issue is the distribution of MW Cepheids across a very large sky area, requiring MW Cepheids to be targeted individually and causing large observational overheads.

Drift scanning observations with \hst\ have allowed the collection of photometry in the SH0ES Wesenheit system for a legacy sample of 75 MW FM Cepheids that feature low extinction and are sufficiently nearby ($\sim 1 - 5$\,kpc) for high-quality \Gaia\ parallaxes \citep{Riess2018phot,Riess2021phot}. \hst\ drift scan observations have further provided parallaxes of 8 Cepheids, with a typical accuarcy of $40-50\,\mu$as \citep{Riess2014,Casertano2016,Riess2018plx}. 

Hundreds of high-quality parallaxes of MW Cepheids have been published by \Gaia\ \citep{Prusti2016gaia,Brown2021gaiaedr3,GaiaDR3summary,GaiaDR3astrometry,Lindegren2021}, an order of magnitude increase over ESA's predecessor {\it Hipparcos} \citep{Hipparcos,VanLeeuwen2007}. Clearly, \Gaia\ has been a \emph{game changer} for DL calibration and has ushered in a new era for SCs. However, \Gaia\ (E)DR3 parallaxes are subject to unfortuante systematics at the level of $10-20\,\mu$as (a $3 - 6\%$ error at $3$\,kpc) that correlate strongly with sky position (ecliptic latitude), magnitude, and color, cf. Lindegren, this volume. \cite{Lindegren2021} (L21) determined parallax offset corrections based on these parameters using bright physical pairs, LMC stars, and quasars. However, several follow-up studies \cite{Zinn2021,Riess2022H0,Molinaro2023} have shown that stars brighter than $G \lesssim 11$\,mag require additional correction at levels similar to the L21 corrections \citep{Groenewegen2023Budapest}. Very recently, \cite{Khan2023plx} showed that the L21 corrections require significant further corrections at bright ($G \lesssim 11$\,mag) magnitudes and that these residual corrections depend significantly on sky position, likely because the bright physical pairs insufficiently sample the full sky. As a result, all SC calibrations in this magnitude range must simultaneously solve for a sample parallax offset, which requires careful analysis and generally comes at the cost of precision, and this notably includes Cepheids that are typically brighter than $11$\,mag.

The best parallax accuracy for MW Cepheids is afforded by the fainter non-variable member (main sequence) stars of open star clusters that host Cepheids \citep{Riess2022clusters,CruzReyes2023}. The numerous member stars improve statistical precision while \Gaia's parallax sysetmatics are best understood in the magnitude probed by the members stars \citep{Lindegren2021,MaizApellaniz2022}. This fortuitous combination yields a typical accuracy of $7\,\mu$arcsec for cluster Cepheids, affording them the same weight as $\sim 9$ field Cepheids for determining $M_{H,1}^W$. Combining cluster and field Cepheids thus has allowed to calibrate $M_{H,1}^W$, and an analogous magnitude based on \Gaia\ photometry, to better than $0.020$\,mag ($0.9\%$ in distance), while simultaneously solving the sample's overall parallax offset to $-13 \pm 5\,\mu$as ($-19 \pm 3\,\mu$as for the larger sample based on \Gaia\ photometry) \citep{CruzReyes2023}.

\subsubsection{Cepheids in the Magellanic Clouds}\label{sec:LMC}

The Magellanic Clouds host very large Cepheid populations that have provided the most important proving grounds for testing stellar models thanks to the ability to observe the Cepheids in great detail at a virtually constant distance. 

LMC Cepheids are particularly valuable due to the extremely accurate distance determined based on 20 detached eclipsing binaries (cf. Pietrzynski \& Graczyk, this volume; \cite{Pietrzynski2019}). The intrinsic depth of the SMC combined with the relatively few (8) detached eclipsing binaries that define its distance \citep{Graczyk2020} has thus far limited the use of the SMC as an anchor galaxy. Photometric observations using \hst's DASH mode have enabled efficient photometric observations of a significant sample of long-period Cepheids, despite the large angular separation among LMC Cepheids \citep{Riess2019}. While ground-based photometry of LMC Cepheids abounds from several surveys \citep{MACHO,Soszynski2019,Macri2015LMC,Ripepi2022VMC}, \hst's photometry provides unmatched photometric consistency and avoids issues related to Earth's atmosphere, notably in the NIR. 

Cepheids in the Magellanic Clouds usefully constrain the slope of the metallicity term, $\gamma$ \cite{Romaniello2022}, since they are less metallic than most Cepheids in the Solar neighborhood, are not heavily extincted, and not subject to \Gaia\ parallax offsets. However, the internal depth of both galaxies must be considered for best results, and correcting for the LMC plane geometry \citep{Ripepi2022VMC} reduces the scatter of the observed Cepheid LL. This effect is even more pronounced in the SMC \citep{Scowcroft2016}, which also allows to extend the metallicity lever to the lowest values \citep{Breuval2022}.

\subsubsection{Cepheids in NGC\,4258}\label{sec:N4258}

The spiral galaxy NGC\,4258 (Messier 106) is a crucial DL anchor thanks to a water megamaser that orbits its supermassive black hole \citep{Humphreys2013}, cf. Braatz \& Pesce this volume. Modeling the observed maser emission as a warped Keplerian disk has allowed to determine the distance to this galaxy to within $1.5\%$ \citep{Reid2019}. Importantly for the DL, this measurement shares virtually no common systematics with \Gaia\ parallaxes, or the modeling of binary star orbits (apart from Kepler's laws) that underlies the LMC distance. The largest population of $669$ Cepheids in NGC\,4258 has been observed in the SH0ES Wesenheit system by \cite{Yuan2022}. However, long-period Cepheids in NGC\,4258 are sufficiently close to detect them using ground-based, 8m-class telescopes \cite{Shapee+2011,Fausnaugh+2015}.

At a distance of $7.6$\,Mpc, Cepheids in NGC\,4258 are observed under conditions that resemble the conditions of other galaxies much more than Cepheids in the MW, or the LMC, notably with regards to the process by which the photometry is collected (cf. Sec.\,\ref{sec:phot}), the typical flux range, the blending of sources, and the metallicity corrections based on HII regions. This makes NGC\,4258 the only anchor galaxy where Cepheids, as well as other SCs, such as the TRGB and J-region AGB methods \citep{Zgirski2021,Madore2022}, can be calibrated in the same way as they are used for determining distances. Thus, NGC\,4258 allows very direct apples-to-apples comparisons for determining distances to individual galaxies, such as M51 \citep{Csoernyei2023M51}. Additionally, the consistent absolute LL calibrations determined using MW parallaxes, the LMC distance, and NGC\,4258 underline the accuracy of the measurements in all three anchor galaxies.

\subsubsection{Cepheids in SN-host galaxies}\label{sec:SNhosts}

The SH0ES DL uses 2150 Cepheids in 37 SN-host galaxies to calibrate the SNIa luminosity zero-point, and these galaxies are located across a wide range of distances, from $6.7$ (M101) to $80$
\,Mpc (NGC\,105), with an average distance of approximately $27$\,Mpc. This dataset represents a major time investment, totaling approximately $1050$ orbits, or the equivalent of $\sim 66$ full days of continuous \hst\ observations. In the nearest galaxies, several fields can be monitored to map Cepheids across galaxies, whereas a single frame (WFC3/NIR footprint is $2.7 \times 2.7$\,arcmin$^2$) centered on the galactic bulge captures the majority of the Cepheids in distant galaxies. 

Cepheids in SN-hosts are discovered at optical wavelengths where their amplitudes are largest, $\sim 1$\,mag. $V-$ (F555W) and $I-$band (F814W) observations are de-phased to estimate mean magnitudes, whereas $H-$band (F160W) observations can usually be used at random phase thanks to the reduced amplitudes. The combination of $H-$band magnitudes, and $V-I$ colors is then used to compute the Wesenheit magnitudes. Further information on Cepheid photometry beyond the Local Group and Wesenheit magnitudes is provided in Sect.\,\ref{sec:phot}. 

Bona fide Cepheids used for the DL are discovered based on their light variations and subsequently vetted using several quality criteria, cf. \cite{Yuan2022}. These cuts include amplitudes, amplitude ratios, dispersion of Cepheid observations around template light curve fits, minimum period of $5$\,d, and average color to ensure a matching temperature range. Finally, the sample is cleaned by $3\sigma$ clipping in all three bands. Impressively, binning the light curves of many Cepheids in period ranges allows to illustrate the Hertzsprung progression, which adds further confidence concerning the equivalence of the distant and nearby Cepheids.

The constant angular resolution of the photometric system corresponds to increasingly larger physical scales, the more distant the SN host galaxy. This creates a fairly clean correlation between the degree to which Cepheids are blended with other sources and the distance at which they are observed. Variations in surface source density, e.g., due to inclination or the region of the galaxy observed, further impact the amount of blending. At large source densities, the term \emph{crowding} refers to heavy blending in fields with many overlapping sources. 

\emph{Crowding corrections} are used to remove background light contributions due to unresolved sources \citep{Ferrarese2000}. The background due to blended sources is measured using artificial star tests that locally insert and remeasure the flux of sources in the vicinity of the target star, and with similar brightness \citep{Riess2011,Riess2016}. 
This is a significant effect in many SN host galaxies (mean of $\sim 0.4$\,mag, up to nearly $1\,$mag at the largest distances), although corrections as small as a few hundredths of a magnitude are also estimated in low density regions of nearby galaxies. The accuracy of crowding corrections has been extensively tested \citep{Riess2022H0} and recently been validated using \jwst's higher spatial resolution \citep{Riess2023crowd}.

\emph{Stellar association bias} arises because wide binaries and open star clusters add flux to Cepheids in NGC\,4258 and SN-host galaxies, while being resolved (and hence not contributing) to Cepheid flux measurements used to calibrate the LL in the LMC and MW \citep{Anderson2018}. Given a typical physical scale of clusters being approximately $4$\,pc, the artificial star tests used for crowding corrections cannot correct for this effect, leading to a small, albeit one-sided bias for the DL because the effect occurs primarily in the SN-host galaxies. Using M31 as a SN-host analog, \cite{Anderson2018} estimated an effect of $\sim 0.007$\,mag per SN-host Cepheid, and a correction of this size has been included in the SH0ES analysis. A similar  estimate has been found in M33 \citep{Breuval2023}, and additional work based on M101 and NGC\,4258 is ongoing (Spetsieri et al., Lengen et al., both in prep.). While a better quantification of this bias will further improve \Ho\ accuracy, it is important to note that stellar association bias cannot resolve the \Ho\ tension, since both the requirements of Cepheid discovery (notably sufficiently large amplitudes and characteristic light curve shapes) and the quality cuts applied to Cepheid candidates (notably based on color and LL $\sigma$ clipping) provide very strong protection against significant  bias. In essence, Cepheids blended by very luminous clusters are either not discovered due to diminished amplitudes, or rejected due to their blue color. 

Metallicity corrections in SN-host galaxies can unfortunately not (yet) be measured directly from Cepheid spectra. Instead, oxygen abundances of Cepheids in SN-host galaxies are estimated based on their galactocentric location and metallicity gradients estimated using HII regions \citep{Bresolin2011}. Care must be taken to ensure the consistency of the derived oxygen abundances, notably with regards to the Solar oxygen abundance problem. In that case, a high degree of agreement between Cepheid oxygen abudances and HII region abundances has been shown \citep[Appendix C]{Riess2022H0}. 

\emph{Relativistic effects} can bias \Ho\ measurements based on the DL because SN-host galaxies are systematically redshifted relative to the anchor galaxies due to cosmic expansion. For example, time dilation leads to systematically overestimated Cepheid periods at greater distances, and thus, to overestimated luminosities and underestimated distances. Correcting the dilated observed Cepheid periods therefore has the effect of \emph{increasing} \Ho, by approximately $0.2\%$ \citep{Anderson2019rlb}. This effect is already accounted for in the SH0ES DL. Two additional relativistic effects have been considered \citep{Anderson2022}: $K-$corrections and the redshifting of the reddening law on Wesenheit magnitudes. While both effects are very small when working with NIR Wesenheit magnitudes, single-band IR observations (e.g., of TRGB stars observed with \jwst) can exceed $1\%$ distance bias at $100$\,Mpc and must therefore be considered as the use of stellar SCs is pushed to ever greater distances.

\subsection{Cepheid Photometry in SN-hosts and NGC\,4258\label{sec:phot}}

The quality of the observed photometry is paramount for accurate luminosity estimates and \Ho. As such, a few words on the measurement process are in order.

Point spread function (PSF) photometry remains the state-of-the-art method for measuring Cepheid light curves, mostly using DAOPHOT and DOLPHOT \citep{DAOPHOT,DOLPHOT}. PSF photometry treats each star as a point source that can be distinguished from neighboring sources and background, and it is used for busy and inhomogeneous scenes, where the simpler aperture photometry approach is not suitable. An important benefit of PSF photometry over aperature photometry is its ability to distinguish resolved and unresolved sources, and PSF fitting produces quantitative information that can be used to study the shape of the light's distribution over the detector. However, the key limitation of PSF photometry is the need to know how light is dispersed by the optical system either a priori, or by measuring the PSF directly from isolated sources. The details of how temporal variations in PSF, e.g., due to telescope breathing or focus changes, affects photometry are challenges where observer experience frequently outperforms automated procedures.

In many ways, measuring photometry \emph{well} is an \emph{art} that requires a significant time investment to master. Given the high hurdle for participating and the prevalence of two methods commonly adopted as black boxes, there is certaintly room for more independent cross-checks. A notable recent study presented new photometric measurements based on the raw \hst\ frames of NGC\,5584 \citep{Javanmardi2021}. Comparison with the photometry by the SH0ES team revealed very good agreement. While further cross-checks would be welcome, it is worth noting that \Ho\ bias would primarily arise due to inconsistency among photometric measurements between near and far galaxies, rather than due to errors involving the absolute flux scale of all galaxies.

\subsection{Baade-Wesselink distances\label{sec:BW}}
The large scale radial pulsations of Cepheids enable distance determination using the ratio of the observed angular ($\Delta \Theta$) and linear radius variations ($\Delta R$), since $D \propto \Delta R / \Delta \Theta$. Such distances are commonly referred to as Baade-Wesselink distances \citep{Baade1926,Wesselink1946}, and their development was intricately linked with establishing pulsations as the cause of Cepheid pulsations \citep{Eddington1917,Lindemann1918,Tiercy1927}. $\Delta \Theta$ is observable for a relatively small number of MW Cepheids using very long baseline optical interferometry \citep{Mourard1997,Kervella1999,Lane2000} as well as surface brightness color relations (SBCRs) \citep{Kervella2004}, which require only photometric measurements and are thus applicable to longer distances. The SpectroPhoto-Interferometry of Pulsating Stars (SPIPS) algorithm performs simultaneous modeling of interferometry, multi-band photometry, and radial velocity observations to determine distance using the same concepts \citep{Merand2015SPIPS,Trahin2021}. $\Delta R = p \cdot \int{v_r d\phi}$ is the product of the integral of the line of sight (radial) velocity variations measured using optical spectra and the projection factor $p$, which translates the observed line of sight velocity to the pulsational velocity as measured from the center of the star. The requirement for optical spectra limits the use of BW distances to relatively nearby objects, such as Cepheids in the MW and the Magellanic Clouds \citep{Storm2011pfac,Storm2011met,Gieren2018}. However, the projection factor $p$ has been a limiting factor to he precision attainable for BW distances due to its complex task of  simultaneously correcting for geometric effects, limb darkening, and dynamical aspects of stellar atmospheres \citep{Nardetto2004,Nardetto2007,Anderson2014,Anderson2016c2c,Anderson2016vlti}. For example, despite the availability of \Gaia\ parallaxes, it has not been possible to determine whether or how $p$ may depend on Cepheid properties, such as pulsation period $P$ \citep{Breitfelder2016,Trahin2021}. While BW distances thus provide a potentially interesting alternative for measuring individual distances to Cepheids, they  do not currently afford the accuracy required to inform \Ho.

\subsection{On the impact of Cepheid multiplicity on \Ho\label{sec:multiple}}
The majority of Cepheids reside in multiple systems \citep{Evans2015crav,Moe2017,Kervella2019pm,Kervella2019pm2}, with estimates of the multiplicity fraction ranging from $60\%$ to more than $100\%$ (many Cepheids occur in triple or even quadruple systems). Cepheid companions can affect distances either astrometrically or photometrically. 

Astrometric issues falsify the parallax measurements of Cepheids and can arise either due to real or apparent orbital motion. The former is caused by the movement around the common center of gravity, whereas apparent orbital motion is a photometric effect in disguise and caused by the shifting photocenter in a binary system containing stars of similar magnitude. Timescales can be used to separate these two effects, with true orbital motion generally being much longer than the pulsation period, and apparent motion occuring in phase with the Cepheid's variability. Poor astrometry thus obviously can affect LL calibration, albeit on a star-by-star basis, increasing scatter rather than biasing LL calibration. Thankfully, radial velocity monitoring can very sensitively estimate whether orbital motion can significantly affect parallax measurements \citep{Anderson2016rvs}. \Gaia's astrometric quality flags, such as RUWE, are also commonly used to identify high quality astrometric solutions.

Photometric distance bias can in principle arise if the calibrated LL does not correspond to the observed LL due to differences in the nature of the companion stars, or the prevalence of multiple star systems. However, these concerns are essentially limited to individual Cepheids \citep{Gallenne2018}, which may stand out significantly from observed LLs. For example, abnormally luminous and red LL outliers in the LMC have been successfully identified as high-value double-lined spectroscopic binaries (SB2s) \citep{Pilecki2021,Pilecki2022}. A similar population of SB2 Cepheids in the MW has not yet been detected, which prompts the interesting question of how much the companion star population of Cepheids may differ from galaxy to galaxy. Reasonable assumptions on the distribution of wide binary separations and contrasts have allowed to constrain a possible DL bias to $< 0.004\%$ \citep{Anderson2018}. 

\subsection{Cepheids in (open) star clusters \label{sec:clusters}}

In the Milky Way, approximately 10\% of Cepheids are associated with (gravitationally bound to) open clusters \citep{CruzReyes2023}. This estimate is based on \Gaia\ (E)DR3 data and considers the 2\,kpc surrounding the Sun, where a rather complete census of both Cepheids and star clusters is currently feasible. Different galaxies present different clustered Cepheid fractions, with the LMC having the highest known fraction. In M31, the fraction of Cepheids in star clusters is much less, approximately 2.5\% \citep{Senchyna2015,Anderson2018}, and M33 has a clustered Cepheid fraction of $3.3\%$ \citep{Breuval2023}. HST UV photometry of M101 shows that Cepheid clustering is highly variable across the two fields in which Cepheids have been observed and that it correlates strongly with star formation (Spetsieri et al. in prep.). 

Dynamical simulations \citep{Dinnbier2022} explain the low fraction of Cepheids that occur in clusters as a consequence of cluster dissociation over time. Cluster dispersal is driven by the violent gas expulsion due to massive star winds at very early times and the dynamical interactions over longer timescales, and mostly occurs well before intermediate-mass stars reach the Cepheid stage. Despite sensitivity to initial conditions, the observed trend that younger Cepheids cluster more frequently strongly supports the paradigm that a majority, if not all, stars are born in clusters \citep{Dinnbier2022formation}. Cepheid clustering is thus both dependent on global and local properties of galaxies, as is observed.

\subsubsection{Circumstellar environments}\label{sec:CSEs}
Circumstellar environments (CSEs) of Cepheids have been detected using mid-IR photometry \citep{Marengo2010,Barmby2011} and HI imaging \citep{Matthews2012}, near- and mid-IR interferometry \citep{Kervella2006CSE,Merand2006CSE,Gallenne2013CSE}, as IR excess among spectral energy distributions \citep{Schmidt2015,Groenewegen2020MWSED,Groenewegen2023SED}, and using {\it Spitzer} spectroscopy \citep{Hocde2020}. Based on the IR excesses, it appears that $H-$band is just short of the region where IR excess becomes most noticeable, and that only a fraction of Cepheids exhibit significant mid-IR excess. At present, it is unclear whether and how CSEs can bias Cepheid distances. However, the following preliminary conclusions appear reasonably secure: a) CSEs can lead to DL bias only if the nature of CSEs differs systematically among anchor and SN-host galaxies \hbox{---} in this regard, CSE distance bias behaves like photometric bias due to companions; b) CSEs may lead individual Cepheids to be outliers from LLs; c) long-wavelength observations, such as \jwst\ NIRCAM/F277W and F444W are likely to be more sensitive to CSEs than $H-$band and shorter wavelength observations; d) if chemical composition significantly affects the presence of CSEs, then metallicity corrections in use today likely already correct the majority of the effect induced by CSEs.

\subsection{Stellar evolution models}\label{sec:theory}
Cepheids pose many interesting challenges to stellar evolution theory, starting from the occurrence of blue loops during core He burning, to observed rates of period change, to predicting the correct instability strip boundaries, to predicting the right luminosity (mass) range where Cepheids occur, to the mass-luminosity relation and its dependence on metallicity, convection, and rotation, and the ability to predict Cepheid variability, among others. Stellar models thus benefit from a large number of observational constraints that allow to understand the physical nature of these important stars and to rigorously test model predictions along several axes at once. With thousands of Cepheids in dozens of galaxies, the DL offers unprecedented means for comparisons, albeit at a reduced level of detail at greater distances. However, given their systematic uncertainties, stellar models have limited predictive power for Cepheid distances. 

\begin{figure}
\centering
\includegraphics[width=1\textwidth]{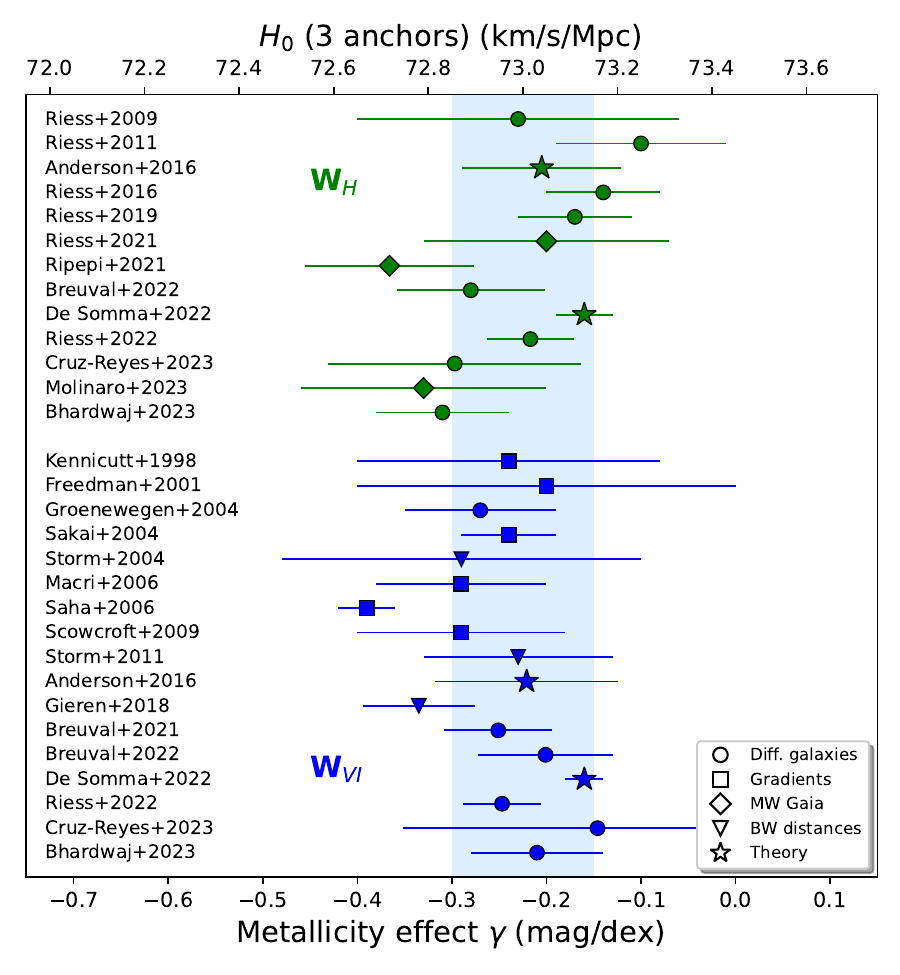}
\caption{The slope of the metallicity effect on Wesenheit magnitudes determined by different methods (bottom x-axis) and the impact of metallicity on \Ho\ (top x-axis). Circles show results based on different galaxies, squares results based on galactic abundance gradients, diamonds results based on MW Cepheids observed with \Gaia, downward triangles results based on BW-type distances, and asterisks results based on stellar evolution models. Figure courtesy of Louise Breuval.\label{fig:metallicity}}
\end{figure}

Areas where stellar models \emph{can} usefully inform the DL include, among others, a) the linearity of the LL (cf. above); b) the dispersion of the LL observed in different passbands; c) differential comparisons of Cepheid luminosity under otherwise identical conditions as required, e.g., to determine the effects of chemical composition on the LL. For example, predictions based on Geneva stellar evolution models \citep{Anderson2016rot} predict a negative sign for the metallicity term, $\gamma$, both in the individual $V-$ and $H-$band PL-relations and in the optical $W_{\mathrm{VI}}$ and near-IR Wesenheit systems $W_{\mathrm{H,VI}}$ used in the SH0ES analysis. Conversely, a widely used set of Cepheid models \citep{Marconi2005,Fiorentino2013,DeSomma2021PA,DeSomma2022} predicts a positive sign for PL-relations in the individual passbands ($\gamma_\lambda > 0$ in those cases) and negatively signed period-Wesenheit relations, that is, when combining a single photometric passband with a color ($\gamma_W < 0$). Recent empirical work by different teams has shown the metallicity effect to have negative sign across several individual photometric bands as well as in the Wesenheit formulations \citep{Gieren2018,Wielgorski2017,Breuval2022,Ripepi2022,Trentin2022,Bhardwaj2023}, cf. Fig.\,\ref{fig:metallicity}. Hence, it appears that the disagreement between the two sets of models is due to different predicted locations of the instability strip boundaries, as well as their dependence on metallicity.

\section{Conclusions\label{sec:conclusions}}
Classical Cepheids provide the base calibration for the \Ho\ measurement using the extragalactic DL. They benefit from more than $110$\,years of study for measuring distances. Recent progress in quantifying systematics has centered on sub-percent effects for \Ho, which ended up sharpening the accuracy of the measurement and the significance of the Hubble tension. In the process, corrections that increase and decrease the value of \Ho\ have been identified. \Gaia\ has revolutionized the calibration of the LL in the MW based on trigonometric parallaxes, and \jwst\ stands to significantly improve \Ho\ precision by better sensitivity and reduced LL dispersion in SN-host galaxies due to lower background (enabled by the $4\times$ better spatial resolution than \hst's WFC3/IR). 

Further improvements of the distance ladder and its experimental setup will yield significant improvements in \Ho\ accuracy, hopefully to deliver the desired $1\%$ direct measurement. Further improvements to \Gaia's parallax systematics are paramount to achieving this level of accuracy. However, despite very significant efforts to identify systematics that could bias \Ho\ sufficiently to explain the \Ho\ constant tension, no such biases have been identified. For this reason, the author is cautiously optimistic that the problem with the \Ho\ constant will eventually enable new interesting insights into the physics governing our Universe.

\begin{acknowledgement}
RIA is funded by the SNSF through an Eccellenza Professorial Fellowship, grant number PCEFP2\_194638 and acknowledges support from the European Research Council (ERC) under the European Union’s Horizon 2020 research and innovation programme (grant agreement No 947660).
\end{acknowledgement}

\bibliographystyle{spphys} 
\bibliography{H0tensionBook_Cepheids} 

\end{document}